%% file: main.tex
\documentclass[10pt,twocolumn,letterpaper]{article}

\usepackage[pagenumbers]{cvpr} 

\usepackage{amsfonts}
\usepackage{amsmath}
\usepackage{booktabs}
\usepackage{multirow}

\input{preamble}
\definecolor{cvprblue}{rgb}{0.21,0.49,0.74}
\usepackage[pagebackref,breaklinks,colorlinks,allcolors=cvprblue]{hyperref}

\def\paperID{} 
\def\confName{CVPR}
\def\confYear{2026}

\title{Tri-Subspaces Disentanglement for Multimodal Sentiment Analysis}

\author{Chunlei Meng\thanks{This study has been Accepted by CVPR 2026.}\thanks{The camera-ready version can be obtained from the official CVPR 2026 website.}\\
Fudan University\\
{\tt\small clmeng23@m.fudan.edu.cn}
\and
Jiabin Luo\\
Peking University\\
\and
Zhenglin Yan\\
Fudan University\\
\and
Zhenyu Yu\\
Fudan University\\
\and
Rong Fu\\
University of Macau\\
\and
Zhongxue Gan\\
Fudan University\\
\and
Chun Ouyang\thanks{Corresponding Author}\\
Fudan University\\
}

\begin{document}
\maketitle
\input{sec/0_abstract}    
\input{sec/1_intro}
\input{sec/2_Related_Work}
\input{sec/3_Proposed_Method}

\input{sec/4_Experiments}

\input{sec/5_Conclusion}

{
    \small
    \bibliographystyle{ieeenat_fullname.bst}
    \bibliography{main}
}

\input{sec/X_suppl}



\end{document}

%% file: sec/0_abstract.tex
\begin{abstract}
Multimodal Sentiment Analysis (MSA) integrates language, visual, and acoustic modalities to infer human sentiment. Most existing methods either focus on globally shared representations or modality-specific features, while overlooking signals that are shared only by certain modality pairs. This limits the expressiveness and discriminative power of multimodal representations. To address this limitation, we propose a Tri-Subspace Disentanglement (TSD) framework that explicitly factorizes features into three complementary subspaces: a common subspace capturing global consistency, submodally-shared subspaces modeling pairwise cross-modal synergies, and private subspaces preserving modality-specific cues. To keep these subspaces pure and independent, we introduce a decoupling supervisor together with structured regularization losses. We further design a Subspace-Aware Cross-Attention (SACA) fusion module that adaptively models and integrates information from the three subspaces to obtain richer and more robust representations. Experiments on CMU-MOSI and CMU-MOSEI demonstrate that TSD achieves state-of-the-art performance across all key metrics, reaching 0.691 MAE on CMU-MOSI and 54.9\% ACC$_7$ on CMU-MOSEI, and also transfers well to multimodal intent recognition tasks. Ablation studies confirm that tri-subspace disentanglement and SACA jointly enhance the modeling of multi-granular cross-modal sentiment cues.
\end{abstract}

%% file: sec/1_intro.tex
\begin{figure}[htbp]
    \centering
    \includegraphics[width=1\linewidth]{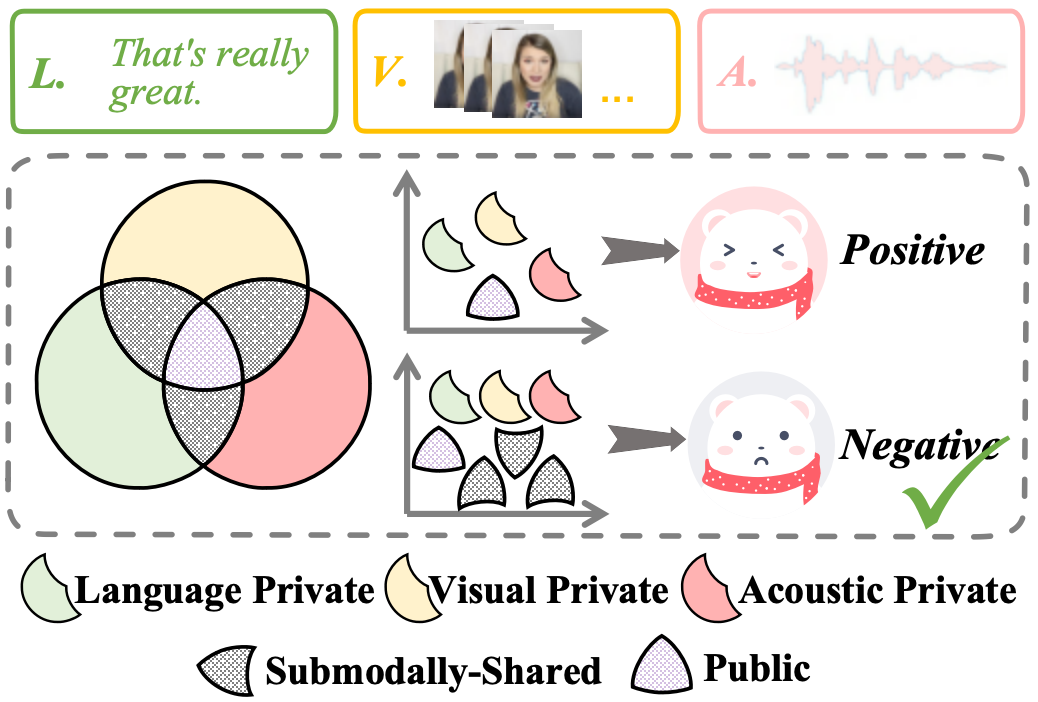}
    \caption{An example of a submodally shared cue: the utterance ``that's really great'' is delivered with a sarcastic tone and a disdainful facial expression. While the lexical content suggests positive sentiment, the conflicting acoustic and visual cues jointly convey negative affect (sarcasm).}
    \label{fig:motivation}
\end{figure}

\begin{figure*}[htbp]
    \centering
    \includegraphics[width=1\linewidth]{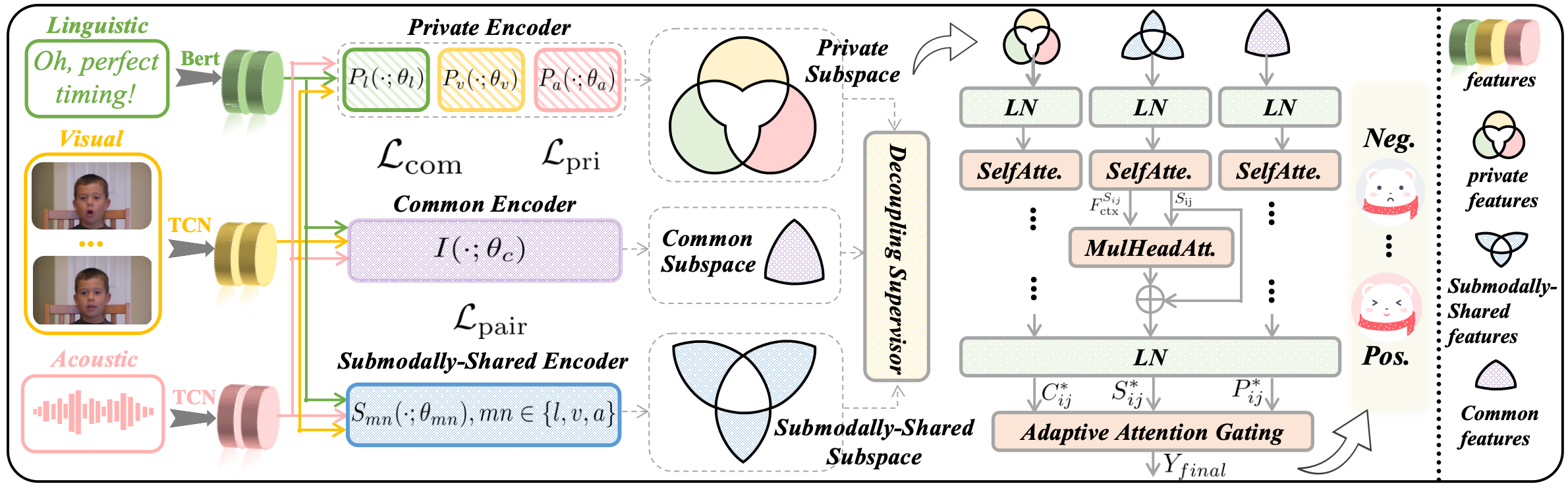}
    \caption{Overview of the proposed TSD framework. Given multimodal inputs, TSD disentangles features into three complementary subspaces and fuses them via a Subspace-Aware Cross-Attention (SACA) module.}
    \label{fig:TSD}
\end{figure*}

\section{Introduction}
\label{sec:intro}

As human-centered AI applications such as empathetic chatbots, social robots, and virtual assistants become increasingly pervasive, accurately modeling human emotions is becoming more critical~\cite{FDMER,CF-ViT}. Consequently, Multimodal Sentiment Analysis (MSA) has attracted growing attention in both academia and industry for its ability to interpret emotions by jointly leveraging acoustic, visual, and linguistic modalities~\cite{ALMT}. However, MSA remains challenging: different modalities follow heterogeneous feature distributions, emotional cues are often sparse or noisy, and informative signals are not always aligned across modalities. These issues call for more effective ways to represent and fuse complementary information across diverse modalities~\cite{Semi-IIN,DLF,cta-net,qiao,rts-vit}.

A major line of work addresses these challenges by factorizing features into common (modality-invariant) and private (modality-specific) subspaces. For example, MISA~\cite{misa} explicitly separates a shared latent space from private spaces capturing each modality’s unique aspects. Follow-up studies such as FDMER~\cite{FDMER}, FDRL~\cite{FDRL}, and DLF~\cite{DLF} refine this binary decomposition to obtain more fine-grained disentangled representations. Other approaches employ self-attention and dynamic fusion to model cross-modal interactions and adaptively combine modalities~\cite{TSDA,d2r,dmd,EMOE}, or incorporate contrastive learning and semi-supervised objectives to better exploit data and auxiliary signals~\cite{confede,CGGM,Semi-IIN,DEVA}. While these methods have advanced MSA performance, they are built on an implicit two-subspace assumption: features are either fully shared across all modalities or entirely private to a single modality, leaving partially shared patterns under-explored.

In practice, many sentiment cues are neither fully global nor strictly modality-specific, but are instead shared across only a subset of modalities. Ignoring these submodally shared signals can lead to incomplete or even misleading sentiment interpretation. As illustrated in Fig.~\ref{fig:motivation}, a speaker may say \textit{``That’s really great''} with a sarcastic tone and a disdainful facial expression. The linguistic content alone suggests positive sentiment, whereas the combination of vocal and facial cues clearly expresses a negative attitude. Here, the true emotion is revealed by the audio--visual pair: these two modalities share a negative sentiment cue (sarcasm) that is not reflected in the language. Prior works~\cite{MInD,TFN} have shown that such intermediate, submodally shared features are semantically meaningful and should be preserved. However, existing common-vs-private frameworks tend to push any signal not present in all modalities into private channels, effectively discarding valuable pairwise or subgroup correlations.

To address this limitation, we propose a \textbf{Tri-Subspace Disentanglement (TSD)} framework that explicitly factorizes the multimodal feature space into three complementary subspaces: (1) a \textbf{common subspace} capturing globally consistent information shared by all modalities, (2) \textbf{submodally shared subspaces} modeling synergies between specific modality pairs, and (3) \textbf{private subspaces} retaining modality-specific characteristics. We introduce a tri-subspace decoupling supervisor together with structured regularization losses to encourage each subspace to remain distinct, i.e., to focus on its designated type of information with minimal leakage. On top of these disentangled representations, we introduce a \textbf{Subspace-Aware Cross-Attention (SACA)} module that dynamically attends to and fuses information from the three subspaces, enabling the model to exploit high-level commonalities, intermediate cross-modal cues, and fine-grained modality-specific details within a unified prediction framework. Experiments on CMU-MOSI, CMU-MOSEI, and a multimodal intent recognition benchmark show that TSD achieves state-of-the-art performance and generalizes well across tasks, while ablation studies further verify the contributions of both tri-subspace disentanglement and SACA fusion. Overall, our contribution lies in an architectural and representation design that explicitly models submodally shared signals, rather than in proposing a new theoretical framework.

%% file: sec/2_Related_Work.tex
\section{Related Work}

\noindent\textbf{Multimodal Sentiment Analysis.}
MSA aims to recognize emotions by leveraging language, visual, and acoustic signals.
Early methods relied on simple feature-level fusion (e.g., concatenation or summation), which struggles to capture complex cross-modal dependencies.
Subsequent works introduced tensor-based models such as TFN~\cite{TFN} and attention mechanisms to better model intra- and inter-modal interactions.
However, MSA remains challenging due to modality-specific noise, temporal misalignment, and semantic inconsistencies across modalities, which can severely degrade fusion quality.

\noindent\textbf{Disentangled Representation Learning.}
To mitigate modality heterogeneity, recent approaches adopt disentangled representations that separate shared and modality-specific components~\cite{MGJR}.
MISA~\cite{misa}, for example, employs a shared-private framework to align modality-invariant features while preserving modality-specific signals, and FDMER~\cite{FDMER} further introduces contrastive objectives to strengthen this separation.
These methods improve robustness and generalization, but typically impose a binary partition (common vs.\ private) and overlook partially shared signals.
In practice, many sentiment cues (e.g., the intensity of anger in speech and facial expressions) are expressed only through specific modality pairs, which cannot be adequately modeled by purely global or purely private subspaces.

\noindent\textbf{Fusion Strategies.}
Fusion design is another key component in MSA.
Beyond early fusion by concatenation or summation, attention-based fusion, modality-specific transformations, and decision-level fusion with gating have been proposed to better capture cross-modal interactions.
Representative methods include EMOE~\cite{EMOE}, which employs emotion-specific experts, DEVA~\cite{DEVA}, which uses auxiliary textual prompts for contextual reasoning, and CMAF~\cite{FDMER}, which performs cross-modal alignment fusion.
Most of these methods, however, still assume that modality contributions are either fully shared or independent, implicitly collapsing partially shared signals into private spaces or enforcing global sharing.
In contrast, our TSD framework explicitly introduces pairwise shared subspaces between modalities and couples them with the SACA fusion module, which performs subspace-level cross-attention and channel re-weighting over common, pairwise shared, and private representations, enabling a more fine-grained modeling of multimodal sentiment cues.

%% file: sec/3_Proposed_Method.tex
\section{Proposed Method}

\subsection{Model Overview}

Given an utterance $U$, our goal is to predict its sentiment polarity by jointly modeling three modalities: linguistic ($l$), visual ($v$), and acoustic ($a$).
Different from previous methods~\cite{EMOE,DLF}, and as illustrated in Fig.~\ref{fig:TSD}, our framework explicitly disentangles multimodal representations into three complementary subspaces:
(i) a fully shared (\emph{common}) subspace that captures global semantic cues shared across all modalities,
(ii) \emph{submodally shared} subspaces that model interactions present only in subsets of modalities, and
(iii) modality-specific (\emph{private}) subspaces that preserve unique and discriminative information for each modality.

Concretely, we first encode $U_l$, $U_v$, and $U_a$ into modality-specific high-level representations and project them into a unified feature space. Dedicated decoupling modules then map these representations into the three subspaces.
The common subspace aggregates modality-invariant information critical for overall affective inference.
The submodally shared subspaces capture associations present in two out of three modalities, enabling the model to exploit pairwise correlations.
The private subspaces retain complementary features unique to each modality, ensuring that unimodal cues are not lost.
Given the representations from all three subspaces, we fuse their outputs and map the fused representation to either a categorical label ($y \in \mathbb{R}^C$) or a continuous sentiment score ($y \in \mathbb{R}$).
The following sections formalize the feature extraction, tri-subspace construction, and their fusion in detail.

\subsection{Feature Representation}

To capture temporal patterns within each modality, we process the visual and acoustic inputs using independent temporal convolutional networks, while language features are extracted by a pre-trained BERT encoder.
For each modality $m \in \{v, a, l\}$, the input sequence is denoted as $\mathbf{X}_m \in \mathbb{R}^{T_m \times d}$, where $T_m$ is the sequence length and $d$ is the feature dimension.
Each modality-specific encoder produces a sequence of hidden representations:
\begin{equation}
\mathbf{H}_m = \text{Encoder}_m(\mathbf{X}_m; \theta^{\text{enc}}_m) \in \mathbb{R}^{T_m \times d_m},
\end{equation}
where $\theta^{\text{enc}}_m$ are the learnable parameters of the encoder for modality $m$.

We then project the encoder outputs into a unified feature space via a linear layer:
\begin{equation}
\mathbf{Z}_m = \text{Proj}_m(\mathbf{H}_m) \in \mathbb{R}^{T_m \times d_z},
\end{equation}
where $\text{Proj}_m$ is a modality-specific fully connected layer and $d_z$ is the shared feature dimension.
These unified representations $\mathbf{Z}_m$ serve as the inputs to the tri-subspace disentanglement module.

\subsection{Tri-Subspaces Disentanglement}
\label{subsec:tri}

Existing multimodal representation learning methods can capture both global consistency and modality-specific information, but often overlook collaborative signals that are shared only by subsets of modalities.
Without explicit modeling, these submodally shared cues may be misattributed to either common or private subspaces, limiting the expressiveness and discriminative power of the learned features.
We therefore introduce a tri-subspace disentanglement framework that factorizes multimodal representations into common, submodally shared, and private subspaces, enabling multi-granular modeling of global consistency, pairwise collaboration, and individual modality characteristics.

Let $\mathcal{M} = \{l,v,a\}$ denote the set of modalities with $|\mathcal{M}| = M = 3$, and let $\mathcal{P}_2(\mathcal{M})$ be the set of all unordered modality pairs.

\subsubsection{Tri-Subspaces Encoders}


\noindent\textbf{Common encoder $I(\cdot; \theta_c)$.}
This encoder extracts modality-invariant, globally consistent features and is shared across all modalities.
It consists of two fully connected layers with GELU activation and LayerNorm.
The output is
\begin{equation}
\mathbf{C}_m = I(\mathbf{Z}_m; \theta_c) \in \mathbb{R}^{T_m \times d_c}.
\end{equation}

\noindent\textbf{Submodally shared encoder $S_{mn}(\cdot; \theta_{mn})$.}
This encoder captures specific correlations between a modality pair $(m,n)$ and extracts the information shared only between the two modalities.
For each unordered modality pair $(m,n) \in \mathcal{P}_2(\{l,v,a\})$, we define a shared-parameter encoder with one fully connected layer followed by a sigmoid activation.
Its outputs are
\begin{equation}
\mathbf{S}_{mn}^{(m)} = S_{mn}(\mathbf{Z}_m; \theta_{mn}), \quad
\mathbf{S}_{mn}^{(n)} = S_{mn}(\mathbf{Z}_n; \theta_{mn}),
\end{equation}
where the superscript $(m)$ indicates that the representation is derived from modality $m$ but lives in the subspace shared with modality $n$.
in practice, we construct $\mathbf{S}{ij}$ by concatenating $\mathbf{S}{ij}^{(i)}$ and $\mathbf{S}_{ij}^{(j)}$ along the temporal dimension.

\noindent\textbf{Private encoder $P_m(\cdot; \theta_m)$.}
For each modality, we also define a private encoder composed of one fully connected layer and a sigmoid activation, which extracts modality-specific representations:
\begin{equation}
\mathbf{P}_m = P_m(\mathbf{Z}_m; \theta_m).
\end{equation}

\subsubsection{Decoupling Supervisor}

To prevent representation mixing across subspaces, we propose a tri-subspace decoupling supervisor.
The supervisor contains three discriminative branches corresponding to the common, submodally shared, and private subspaces, respectively.
Each branch is a two-layer perceptron with GELU activation~\cite{GELU}, taking the corresponding subspace embedding as input.
During training, embeddings from the three subspaces are pooled over time and mixed within a mini-batch, and the supervisor predicts the true source subspace of each embedding.

Let $\rho(\cdot)$ be a temporal pooling operator (e.g., mean pooling) that maps a token sequence to an utterance-level vector.
We define
\begin{equation}
\mathbf{c}_m = \rho(\mathbf{C}_m), \quad
\mathbf{s}_{mn}^{(m)} = \rho(\mathbf{S}_{mn}^{(m)}), \quad
\mathbf{p}_m = \rho(\mathbf{P}_m).
\end{equation}
Let $D_{\mathrm{com}}$, $D_{\mathrm{sub}}$, and $D_{\mathrm{pri}}$ denote the probabilities assigned by the supervisor to the ``common'', ``submodally shared'', and ``private'' labels, respectively.
The supervision loss is defined as
\begin{equation}
\begin{aligned}
\mathcal{L}_{\mathrm{sup}} =\;
& -\frac{1}{M} \sum_{m=1}^{M} \Bigg[
        \mathbb{E}_{x \sim \mathcal{D}}\!\left[\log D_{\mathrm{com}}(\mathbf{c}_m)\right] \\
& \quad + \sum_{n \neq m} \mathbb{E}_{x \sim \mathcal{D}}\!\left[\log D_{\mathrm{sub}}\!\big(\mathbf{s}_{mn}^{(m)}\big)\right] \\
& \quad + \mathbb{E}_{x \sim \mathcal{D}}\!\left[\log D_{\mathrm{pri}}(\mathbf{p}_m)\right]
    \Bigg],
\end{aligned}
\end{equation}
where $\mathcal{D}$ denotes the data distribution (in practice, expectations are approximated by mini-batch averages).
Backpropagation through this loss explicitly enforces clear boundaries among the three subspaces and suppresses information leakage.
Compared with the modality discriminators used in FDMER and related methods, our supervisor jointly discriminates between three subspaces, enforcing stricter disentanglement beyond the shared private dichotomy.

\subsubsection{Subspace Loss Functions}

To further regularize each subspace and capture multi-granular relationships, we design a set of subspace-specific losses.
All losses are computed on utterance-level representations obtained via $\rho(\cdot)$.

\noindent\textbf{Common consistency loss.}
For the common subspace, we adopt a consistency loss $\mathcal{L}_{\mathrm{com}}$ that encourages alignment of modality-invariant features and reduces distribution gaps across modalities:
\begin{equation}
\mathcal{L}_{\mathrm{com}}
= \frac{1}{|\mathcal{P}_2(\mathcal{M})|} \sum_{(m,n)\in \mathcal{P}_2(\mathcal{M})}
   \big\| \mathbf{c}_m - \mathbf{c}_n \big\|_2^2.
\end{equation}

\noindent\textbf{Pairwise collaboration loss.}
To model the pairwise collaboration between modalities, we introduce a pairwise consistency loss $\mathcal{L}_{\mathrm{pair}}$ that encourages submodally shared representations from the two directions to encode only joint information of each modality pair while suppressing modality-specific noise:
\begin{equation}
\mathcal{L}_{\mathrm{pair}}
= \frac{1}{|\mathcal{P}_2(\mathcal{M})|} \sum_{(m,n)\in \mathcal{P}_2(\mathcal{M})}
   \big\| \mathbf{s}_{mn}^{(m)} - \mathbf{s}_{mn}^{(n)} \big\|_2^2.
\end{equation}

\noindent\textbf{Private disparity loss.}
For the private subspaces, we employ the Hilbert--Schmidt Independence Criterion (HSIC)~\cite{HSIC} to encourage independence between private representations of different modalities and prevent contamination from shared information.
Given two private feature sets $\mathbf{p}_{m_1}$ and $\mathbf{p}_{m_2}$ from a mini-batch of size $n$, HSIC is defined as
\begin{equation}
\mathrm{HSIC}(\mathbf{p}_{m_1}, \mathbf{p}_{m_2})
= (n-1)^{-2} \, \mathrm{Tr}(U K_{m_1} U K_{m_2}),
\end{equation}
where $K_{m_1}$ and $K_{m_2}$ are the $n \times n$ Gram matrices of private features under a linear kernel,
$U = I - \tfrac{1}{n} \mathbf{e}\mathbf{e}^\top$,
$I$ is the identity matrix, and $\mathbf{e}$ is an all-ones vector.
The overall private disparity loss aggregates HSIC over all modality pairs:
\begin{equation}
\mathcal{L}_{\mathrm{pri}}
= \frac{1}{|\mathcal{P}_2(\mathcal{M})|}
  \sum_{(m_1, m_2)\in \mathcal{P}_2(\mathcal{M})}
   \mathrm{HSIC}(\mathbf{p}_{m_1}, \mathbf{p}_{m_2}).
\end{equation}

\noindent\textbf{Orthogonality loss.}
To further disentangle the three subspaces within each modality, we introduce an orthogonality constraint $\mathcal{L}_{\mathrm{ort}}$ that penalizes overlap among common, submodally shared, and private representations:
\begin{equation}
\begin{aligned}
\mathcal{L}_{\mathrm{ort}} =\;
& \frac{1}{M} \sum_{m \in \mathcal{M}} \Big[
    \big\|\mathbf{C}_m^\top \mathbf{P}_m\big\|_F^2 \\
& \quad + \sum_{n \neq m} \big(
        \big\|\mathbf{S}_{mn}^{(m)\top} \mathbf{P}_m\big\|_F^2
      + \big\|\mathbf{S}_{mn}^{(m)\top} \mathbf{C}_m\big\|_F^2
      \big)
\Big],
\end{aligned}
\end{equation}
where $\|\cdot\|_F$ denotes the Frobenius norm.

\noindent\textbf{Tri-subspace training objective.}
The overall objective for tri-subspace disentanglement combines all the above losses:
\begin{equation}
\mathcal{L}_{\mathrm{TS}}
= \mathcal{L}_{\mathrm{com}}
+ \lambda_{1} \mathcal{L}_{\mathrm{pair}}
+ \lambda_{2} \mathcal{L}_{\mathrm{pri}}
+ \lambda_{3} \mathcal{L}_{\mathrm{ort}}
+ \lambda_{4} \mathcal{L}_{\mathrm{sup}},
\end{equation}
where $\lambda_{1\text{--}4}$ are hyperparameters controlling the contribution of each regularization term.

\subsection{Subspace-Aware Cross-Attention Fusion}
\label{subsec:saca}

The three types of feature subspaces (common, submodally shared, and private) capture complementary aspects of multimodal information, and their contributions to the final decision are generally unequal.
Simple feature concatenation ignores these structural differences, leading to redundancy and loss of discriminative cues.
To address this, we propose a Subspace-Aware Cross-Attention (SACA) module that performs context-aware interaction and adaptive fusion among all subspaces.

Let $\mathbf{C}_m$, $\mathbf{S}_{ij}^{(m)}$, and $\mathbf{P}_m$ be the tri-subspace representations from Sec.~\ref{subsec:tri}.
We first apply self-attention for subspace-wise refinement:
\begin{equation}
\left\{
\begin{aligned}
\tilde{\mathbf{C}}_m &= \mathrm{SelfAtt}\!\big(\mathrm{LN}(\mathbf{C}_m)\big),
   && m \in \mathcal{M}, \\
\tilde{\mathbf{S}}_{ij} &= \mathrm{SelfAtt}\!\big(\mathrm{LN}(\mathbf{S}_{ij})\big),
   && (i,j) \in \mathcal{P}_2(\mathcal{M}), \\
\tilde{\mathbf{P}}_m &= \mathrm{SelfAtt}\!\big(\mathrm{LN}(\mathbf{P}_m)\big),
   && m \in \mathcal{M},
\end{aligned}
\right.
\end{equation}
where $\mathbf{S}_{ij}$ denotes the subspace shared between modalities $i$ and $j$; in practice, we construct $\mathbf{S}_{ij}$ by concatenating $\mathbf{S}_{ij}^{(i)}$ and $\mathbf{S}_{ij}^{(j)}$ along the temporal dimension, and $\mathrm{LN}$ is layer normalization.

\noindent\textbf{Context construction for cross-subspace attention.}
For each subspace, we dynamically construct a context set that collects the most relevant complementary signals:
\begin{equation}
\begin{aligned}
F^{\tilde{\mathbf{S}}_{ij}}_{\mathrm{ctx}} &=
\big[
    \tilde{\mathbf{S}}_{ij};\ \tilde{\mathbf{C}}_i;\ \tilde{\mathbf{C}}_j;\ \tilde{\mathbf{P}}_i;\ \tilde{\mathbf{P}}_j
\big], \\
F^{\tilde{\mathbf{C}}_m}_{\mathrm{ctx}} &=
\big[
    \tilde{\mathbf{C}}_m;\ \{\tilde{\mathbf{S}}_{mj} \mid j \neq m\};\ \tilde{\mathbf{P}}_m
\big], \\
F^{\tilde{\mathbf{P}}_m}_{\mathrm{ctx}} &=
\big[
    \tilde{\mathbf{P}}_m;\ \{\tilde{\mathbf{S}}_{mj} \mid j \neq m\};\ \tilde{\mathbf{C}}_m
\big],
\end{aligned}
\end{equation}
where $[\,\cdot\,]$ denotes concatenation along the temporal (token) dimension.

\noindent\textbf{Cross-subspace attention enhancement.}
Taking the submodally shared subspace $\tilde{\mathbf{S}}_{ij}$ as an example, we define
\begin{equation}
Q = W_Q \tilde{\mathbf{S}}_{ij}, \quad
K = W_K F^{\tilde{\mathbf{S}}_{ij}}_{\mathrm{ctx}}, \quad
V = W_V F^{\tilde{\mathbf{S}}_{ij}}_{\mathrm{ctx}},
\end{equation}
where $W_Q$, $W_K$, and $W_V$ are learnable projection matrices.
The cross-attention output is computed as
\begin{equation}
\mathbf{O} = \mathrm{softmax}\!\left( \frac{Q K^\top}{\sqrt{d_k}} \right) V,
\end{equation}
where $d_k$ is the key dimension.
For richer modeling, we use multi-head attention and obtain the enhanced subspace representation via a residual connection:
\begin{equation}
\mathbf{S}_{ij}^*
= \mathrm{LN}\big(
    \tilde{\mathbf{S}}_{ij}
    + \mathrm{MultiHeadAttn}(\tilde{\mathbf{S}}_{ij},
    F^{\tilde{\mathbf{S}}_{ij}}_{\mathrm{ctx}},
    F^{\tilde{\mathbf{S}}_{ij}}_{\mathrm{ctx}})
  \big),
\end{equation}
where $\mathrm{MultiHeadAttn}$ denotes the standard multi-head attention operator.
The same procedure is applied to the common and private subspaces to obtain $\mathbf{C}_m^*$ and $\mathbf{P}_m^*$.
The residual connections preserve the original information of each subspace while injecting cross-subspace context.

\noindent\textbf{Hierarchical gated fusion.}
All context-enhanced subspaces are then aggregated:
\begin{equation}
F_{\mathcal{S}} =
\big\{
\mathbf{C}_l^*, \mathbf{C}_a^*, \mathbf{C}_v^*,\,
\mathbf{S}_{la}^*, \mathbf{S}_{lv}^*, \mathbf{S}_{av}^*,\,
\mathbf{P}_l^*, \mathbf{P}_a^*, \mathbf{P}_v^*
\big\}.
\end{equation}
A gating network computes the adaptive weight $\psi_k$ for each subspace:
\begin{equation}
\psi_k = \frac{\exp\big(g_k(F_{\mathcal{S}})\big)}{\sum_{k' \in \mathcal{K}} \exp\big(g_{k'}(F_{\mathcal{S}})\big)},
\end{equation}
where $g_k(\cdot)$ is a learnable scoring function and $\mathcal{K}$ indexes all subspaces (here $|\mathcal{K}| = 9$).
The final multimodal representation is obtained as a weighted sum:
\begin{equation}
\mathbf{Y}_{\mathrm{final}} = \sum_{k \in \mathcal{K}} \psi_k \cdot F_{\mathcal{S}}^{(k)},
\end{equation}
where $F_{\mathcal{S}}^{(k)}$ denotes the enhanced output of the $k$-th subspace.
Finally, $\mathbf{Y}_{\mathrm{final}}$ is fed into fully connected layers for sentiment prediction.

\subsection{Objective Optimization}

For classification tasks, we use the cross-entropy loss, while for regression tasks the mean squared error serves as the task loss $\mathcal{L}_{\text{task}}$.
The final optimization objective combines the task loss with the tri-subspace regularization terms:
\begin{equation}
\mathcal{L}_{\text{all}} =
\mathcal{L}_{\text{task}}
+ \mathcal{L}_{\text{TS}}.
\end{equation}

\begin{table*}[ht]
\caption{Performance comparison on CMU-MOSI and CMU-MOSEI. Each cell reports results under aligned/unaligned settings as $a / b$ ($a$: aligned, $b$: unaligned). *: values from EMOE~\cite{EMOE}. The last row lists the $\pm$ standard deviation of TSD over $n=5$ random seeds.}
\centering
\resizebox{\textwidth}{!}{%
\setlength{\tabcolsep}{2.6pt} 
\renewcommand{\arraystretch}{1.0} 

\begin{tabular}{c||cccc||cccc}

\toprule
\multirow{2}{*}{\textbf{Methods}}
& \multicolumn{4}{c}{\textbf{CMU-MOSI}}
& \multicolumn{4}{c}{\textbf{CMU-MOSEI}} \\
& MAE($\downarrow$) & ACC$_7$(\%) & ACC$_2$(\%) & F1(\%)
& MAE($\downarrow$) & ACC$_7$(\%) & ACC$_2$(\%) & F1(\%) \\
\midrule
\midrule
EF-LSTM*      & 1.386 / 1.420 & 33.7 / 31.0 & 75.3 / 73.6 & 75.2 / 74.5
              & 0.620 / 0.594 & 47.4 / 46.3 & 78.2 / 76.1 & 77.9 / 75.9 \\
TFN*~\cite{TFN}          & 0.953 / 0.995 & 31.9 / 35.3 & 78.8 / 76.5 & 78.9 / 76.6
              & 0.574 / 0.573 & 50.9 / 50.2 & 80.4 / 84.2 & 80.7 / 84.0 \\
LMF*~\cite{LMF}          & 0.931 / 0.963 & 36.9 / 31.1 & 78.7 / 79.1 & 78.7 / 79.1
              & 0.564 / 0.565 & 52.3 / 51.9 & 84.7 / 83.8 & 84.5 / 83.9 \\
MFN*~\cite{EMOE}          & 0.964 / 0.971 & 35.6 / 34.7 & 78.4 / 80.0 & 78.4 / 80.1
              & 0.574 / 0.567 & 50.8 / 51.3 & 84.0 / 83.2 & 84.0 / 83.3 \\
MuLT~\cite{MuLT}          & 0.936 / 0.933 & 35.1 / 33.2 & 80.0 / 80.3 & 80.1 / 80.3
              & 0.572 / 0.556 & 52.3 / 53.2 & 82.7 / 84.0 & 82.8 / 84.0 \\
MISA~\cite{misa}          & 0.754 / 0.742 & 41.8 / 43.6 & 84.2 / 83.8 & 84.2 / 83.9
              & 0.543 / 0.557 & 52.3 / 51.0 & 85.3 / 84.8 & 85.1 / 84.8 \\
FDMER~\cite{FDMER}         & -     / 0.725 & -   /  44.2 & -    / 84.6 & -    / 84.7
              & -     / 0.536 & -   /  53.8 & -   / 84.1 &   -   / 84.0 \\
ConFEDE~\cite{confede}       & -     / 0.742 & -   /  46.3 & -    / 84.2 & -    / 84.2
              & -     / 0.523 & -   /  54.9 & -   / 81.8 &   -   / 82.3 \\
Self-MM*~\cite{self-mm}     & 0.738 / 0.724 & 45.3 / 45.7 & 84.9 / 83.4 & 84.9 / 83.6
              & 0.540 / 0.535 & 53.2 / 52.9 & 84.5 / 85.3 & 84.3 / 84.8 \\
DMD*~\cite{dmd}         & 0.721 / 0.721 & 46.2 / 46.7 & 83.2 / 84.0 & 83.2 / 84.0
              & 0.546 / 0.536 & 52.4 / 53.1 & 84.8 / 84.7 & 84.7 / 84.7 \\
DEVA~\cite{DEVA}          & -     / 0.730 & -   /  46.3 & -    / 84.4 & -    / 84.5
              & -     / 0.541 & -   /  52.3 & -   / 83.3 &  -    / 82.9 \\
DLF~\cite{DLF}           & -     / 0.731 & -   /  47.1 & -    / 85.1 & -    / 85.1
              & -     / 0.536 & -   /  53.9 & -   / 84.4 &  -    / 85.3 \\
EMOE*~\cite{EMOE}  & 0.710 / 0.697 & 47.7 / 47.8 & 85.4 / 85.4 & 85.4 / 85.3
              & 0.536 / 0.530 & 54.1 / 53.9 & 85.3 / 85.5 & 85.3 / 85.5 \\

\midrule
\textbf{TSD (Ours)}  & \textbf{0.701} / \textbf{0.691} & \textbf{48.8} / \textbf{49.0} & \textbf{86.3} / \textbf{86.5} & \textbf{86.3} / \textbf{86.6}
              & \textbf{0.529} / \textbf{0.525} & \textbf{54.9} / \textbf{54.6} & \textbf{85.8} / \textbf{86.2} & \textbf{85.9} / \textbf{86.2} \\

\textbf{$\pm$Std. Dev.} & $\pm0.005$ / $\pm0.005$ & $\pm0.3$/$\pm0.3$ & $\pm0.1$/$\pm0.1$ & $\pm0.3$/$\pm0.3$
                & $\pm0.005$ / $\pm0.006$ & $\pm0.2$/$\pm0.2$ & $\pm0.1$/$\pm0.1$ & $\pm0.4$/$\pm0.4$ \\

\bottomrule
\end{tabular}
} 

\label{tab:main}
\end{table*}

%% file: sec/4_Experiments.tex
\section{Experiments}

\subsection{Experimental Settings}

\textbf{Datasets.}
We evaluate TSD on three widely used multimodal benchmarks.
CMU-MOSI~\cite{Cmu-mosi} contains 2{,}199 opinion video segments with aligned acoustic and visual features, split into 1{,}284 training, 229 validation, and 686 test samples.
CMU-MOSEI~\cite{Cmu-mosei} includes 22{,}856 segments, with 16{,}326 for training, 1{,}871 for validation, and 4{,}659 for testing.
Both datasets provide sentiment scores in $[-3,3]$, where $-3$ and $3$ denote strongly negative and strongly positive, respectively.
MIntRec~\cite{MIntRec} is a Multimodal Intent Recognition (MIR) benchmark with 2{,}224 samples from 20 intent categories and standard train/validation/test splits.

\textbf{Evaluation Metrics.}
Following prior work~\cite{EMOE}, we evaluate MSA on CMU-MOSI and CMU-MOSEI with Acc-2, Acc-7, F1-score, and Mean Absolute Error (MAE).
For MIR on MIntRec, we report Accuracy, F1-score, Precision, and Recall.

\textbf{Implementation Details.}
All models are implemented in PyTorch and trained on an NVIDIA A100 GPU.
We use a batch size of 16, a weight decay of $1 \times 10^{-5}$, and the Adam optimizer, and train up to 50 epochs with early stopping on validation performance.
To reduce randomness, each method is run with $n=5$ different random seeds, and we report averaged results.

\subsection{Comparison with State-of-the-Art Methods}

We compare TSD with a broad range of representative baselines, including early fusion methods, attention-based fusion models, and recent disentanglement-based approaches (we provide a Model Zoo, details in the Appendix).

\textbf{Results on MSA benchmarks.}
We follow the standard practice and evaluate all methods under \emph{aligned} and \emph{unaligned} settings.
The aligned setting assumes perfect temporal synchronization across modalities, while the unaligned setting simulates more realistic asynchrony.
In all tables, we report results as $a/b$, where $a$ is the aligned score and $b$ is the unaligned score.

Table~\ref{tab:main} summarizes the results on CMU-MOSI and CMU-MOSEI.
TSD consistently achieves the best performance across all metrics and both settings.
On CMU-MOSI (aligned), TSD reaches an MAE of $0.703$ and an Acc-2 of $85.7\%$, outperforming the strongest baseline EMOE~\cite{EMOE} by $0.007$ MAE and $0.3\%$ Acc-2.
Under the unaligned setting, TSD further improves MAE to $0.691$ and Acc-2 to $86.5\%$, showing strong robustness to temporal perturbations.
We also observe consistent gains in Acc-7 and F1-score, indicating better fine-grained sentiment modeling. On CMU-MOSEI, TSD exhibits similar advantages.
In the unaligned setting, TSD achieves an MAE of $0.535$ and an Acc-2 of $85.6\%$, surpassing EMOE~\cite{EMOE} by $0.001$ MAE and $0.1\%$ Acc-2 while also improving Acc-7 and F1.
These improvements demonstrate that simply separating shared and private spaces is not sufficient: explicitly modeling common, submodally shared, and private subspaces, together with SACA fusion, yields more expressive and robust multimodal sentiment representations.

\textbf{MIR benchmark results.}
Table~\ref{tab:MIR} reports results on the MIntRec MIR task.
TSD outperforms strong baselines on all metrics.
In particular, it achieves 73.67\% Accuracy and 71.76\% F1, clearly outperforming EMOE~\cite{EMOE} and other recent methods such as CAGC~\cite{CAGC} and GsiT~\cite{GsiT}.
This confirms that the proposed tri-subspace architecture generalizes beyond sentiment regression to broader multimodal classification tasks.

To verify that improvements are not due to randomness, we perform seed-wise paired $t$-tests ($n=5$, Holm-corrected) under both aligned and unaligned settings.
The results show that TSD’s gains are statistically significant across most metrics and datasets (see Appendix for detailed statistics).

\begin{table}[t]
  \caption{Performance comparison on the MIntRec dataset (\%).}
  \centering
  \setlength{\tabcolsep}{5pt}
  \renewcommand{\arraystretch}{0.9}
  \begin{tabular}{lcccc}
    \toprule
    \textbf{Method} & \textbf{Accuracy} & \textbf{F1} & \textbf{Precision} & \textbf{Recall} \\
    \midrule
    MAG-BERT~\cite{MAG-BERT} & 70.34 & 68.19 & 68.31 & 69.36 \\
    MMIM~\cite{GsiT}         & 71.21 & 68.70 & 69.20 & 68.90 \\
    MuLT~\cite{MuLT}         & 72.58 & 69.36 & 70.73 & 69.47 \\
    MISA~\cite{misa}         & 72.36 & 70.57 & 71.24 & 70.41 \\
    CAGC~\cite{CAGC}         & 73.03 & 70.62 & 70.86 & 70.55 \\
    EMOE~\cite{EMOE}         & 72.58 & 70.73 & 72.08 & 70.86 \\
    GsiT~\cite{GsiT}         & 72.60 & 69.40 & 69.40 & 70.10 \\
    \textbf{TSD (Ours)}      & \textbf{73.67} & \textbf{71.76} & \textbf{72.86} & \textbf{71.90} \\
    \bottomrule
  \end{tabular}
  \label{tab:MIR}
\end{table}

\begin{table}[t]
\caption{Ablation studies of TSD on the three benchmarks. For MOSI and MOSEI we report MAE ($\downarrow$) and ACC$_7$ (\%); for MIntRec we report ACC (\%). $\dagger$ denotes the aligned setting.}
\centering
\setlength{\tabcolsep}{3pt}
\renewcommand{\arraystretch}{1.0}
\begin{tabular}{l||cc||cc|c}
\toprule
\multirow{2}{*}{\textbf{Model}}
    & \multicolumn{2}{c||}{\textbf{MOSI}}
    & \multicolumn{2}{c|}{\textbf{MOSEI}}
    & \textbf{MIntRec} \\
    & MAE & ACC$_7$ & MAE & ACC$_7$ & ACC \\
\midrule
\midrule
\textbf{TSD$^\dagger$ (Ours)} & \textbf{0.703} & \textbf{47.9} & \textbf{0.535} & \textbf{54.3} & -- \\
\textbf{TSD (Ours)}           & \textbf{0.691} & \textbf{48.0} & \textbf{0.530} & \textbf{54.2} & \textbf{73.67} \\
\midrule
\multicolumn{6}{c}{\textit{Importance of Modality}} \\
w/o Linguistic   & 1.010 & 35.5 & 0.830 & 39.8 & 54.3 \\
w/o Acoustic     & 0.810 & 45.0 & 0.620 & 51.2 & 69.3 \\
w/o Visual       & 0.850 & 44.1 & 0.650 & 50.0 & 67.8 \\
\midrule
\multicolumn{6}{c}{\textit{Importance of Representations}} \\
w/o Common       & 0.713 & 46.7 & 0.545 & 53.2 & 72.0 \\
w/o Private      & 0.719 & 46.5 & 0.547 & 53.0 & 71.7 \\
w/o Sub-Shared   & 0.705 & 46.9 & 0.543 & 53.4 & 72.3 \\
Non-Disen.       & 0.722 & 46.1 & 0.550 & 52.6 & 71.1 \\
\midrule
\multicolumn{6}{c}{\textit{Different Fusion Mechanisms}} \\
Sum$^\dagger$      & 0.724 & 47.6 & 0.542 & 53.8 & 73.1 \\
Sum                & 0.713 & 47.7 & 0.537 & 54.0 & 73.3 \\
Concat$^\dagger$   & 0.727 & 47.2 & 0.544 & 53.7 & 72.7 \\
Concat             & 0.715 & 47.4 & 0.538 & 53.9 & 73.0 \\
CMAF~\cite{FDMER}  & 0.710 & 47.5 & 0.533 & 53.8 & 73.0 \\
\midrule
\multicolumn{6}{c}{\textit{Importance of Regularization}} \\
w/o $\mathcal{L}_{\mathrm{com}}$ & 0.719 & 46.7 & 0.546 & 53.0 & 71.9 \\
w/o $\mathcal{L}_{\mathrm{pair}}$& 0.710 & 47.0 & 0.539 & 53.5 & 72.4 \\
w/o $\mathcal{L}_{\mathrm{pri}}$ & 0.712 & 47.2 & 0.541 & 53.6 & 72.6 \\
w/o $\mathcal{L}_{\mathrm{ort}}$ & 0.708 & 47.4 & 0.537 & 53.7 & 72.9 \\
w/o $\mathcal{L}_{\mathrm{sup}}$ & 0.721 & 46.9 & 0.548 & 52.8 & 71.9 \\
Only Task Loss                    & 0.735 & 45.8 & 0.555 & 52.1 & 70.6 \\
\bottomrule
\end{tabular}
\label{tab:ablation-2}
\end{table}

\subsection{Ablation Studies}
\label{subsec:ablation}

We conduct ablations along four axes: modality contribution, representation spaces, fusion mechanisms, and regularization design (Table~\ref{tab:ablation-2}).

\textbf{Importance of modality.}
Removing any modality causes a clear performance drop.
Excluding the linguistic modality leads to the largest degradation (e.g., MAE $1.010$ and ACC$_7$ $35.5$ on MOSI), confirming its central role in sentiment and intent understanding.
Dropping acoustic or visual inputs leads to milder but still substantial declines on all three benchmarks, showing that prosody and facial expressions provide complementary cues and that three-way fusion is beneficial for robust performance.

\textbf{Importance of representations.}
We then ablate the common (global), private (modality-specific), and sub-shared (submodally shared) subspaces by removing their outputs from the fusion and prediction heads while keeping the encoders unchanged.
Removing the common subspace harms performance across all benchmarks, indicating the importance of modality-invariant global semantics.
Removing private subspaces also degrades performance, reflecting the need to preserve unimodal nuances.
Removing the sub-shared subspace further reduces accuracy compared with the full model, especially on MIntRec, highlighting that pairwise shared signals still contribute beyond purely global and purely private spaces.
The Non-Disen. variant bypasses disentanglement and directly fuses features; although it remains close to some single-subspace variants, it consistently underperforms the full tri-subspace design, supporting the effectiveness of tri-subspace factorization.

\textbf{Different fusion mechanisms.} We compare SACA with several common fusion strategies: simple summation (Sum), concatenation (Concat), and the cross-modal alignment fusion CMAF~\cite{FDMER}, evaluating both aligned and unaligned variants for Sum and Concat.
SACA consistently outperforms these baselines.
For example, on MOSI (aligned), replacing SACA with Sum leads to higher MAE and lower ACC$_7$, and similar degradations are observed with Concat.
Compared with CMAF, SACA also achieves lower MAE and higher ACC$_7$ on both MOSI and MOSEI, with consistent gains on MIntRec.
These results indicate that SACA provides a more adaptive and discriminative integration of tri-subspace representations.

\textbf{Importance of regularization.} We study the effect of each regularization term.
Removing any structural loss results in noticeable degradation.
Dropping the common consistency loss $\mathcal{L}_{\mathrm{com}}$ or the decoupling supervisor loss $\mathcal{L}_{\mathrm{sup}}$ leads to the largest declines, showing their importance for global alignment and subspace purity.
Removing pairwise consistency $\mathcal{L}_{\mathrm{pair}}$, private disparity $\mathcal{L}_{\mathrm{pri}}$, or orthogonality $\mathcal{L}_{\mathrm{ort}}$ also harms performance, reflecting their roles in modeling pairwise interactions, preventing information leakage, and reducing redundancy.
When all structural regularizers are removed and only the task loss is used (Only Task Loss), TSD degenerates into a generic fusion model and achieves the lowest performance on all three benchmarks, confirming the necessity of our regularization design.

\subsection{Further Analysis}

\textbf{Qualitative analysis.}
To qualitatively examine the effect of submodally shared subspaces, we visualize prediction trajectories on CMU-MOSI.
For utterances such as \emph{``Oh, perfect timing!''}, where the lexical content alone appears positive or nearly neutral but sarcasm is conveyed by tone and facial expression, variants that only use common and private subspaces tend to underestimate the negativity.
In the w/o Sub-Shared variant, audio--visual cues are either over-smoothed in the common subspace or confined to private subspaces with limited influence on the final prediction, leading to clear deviations from the ground-truth score.
With explicit sub-shared subspaces, TSD can preserve these cross-modal interactions and allow SACA to emphasize them when they are informative, yielding trajectories that follow the reference more closely. Similar behavior is observed in examples such as \emph{``Great, just what I needed.''}, where non-verbal signals dominate.

\begin{figure}[t]
    \centering
    \includegraphics[width=\linewidth]{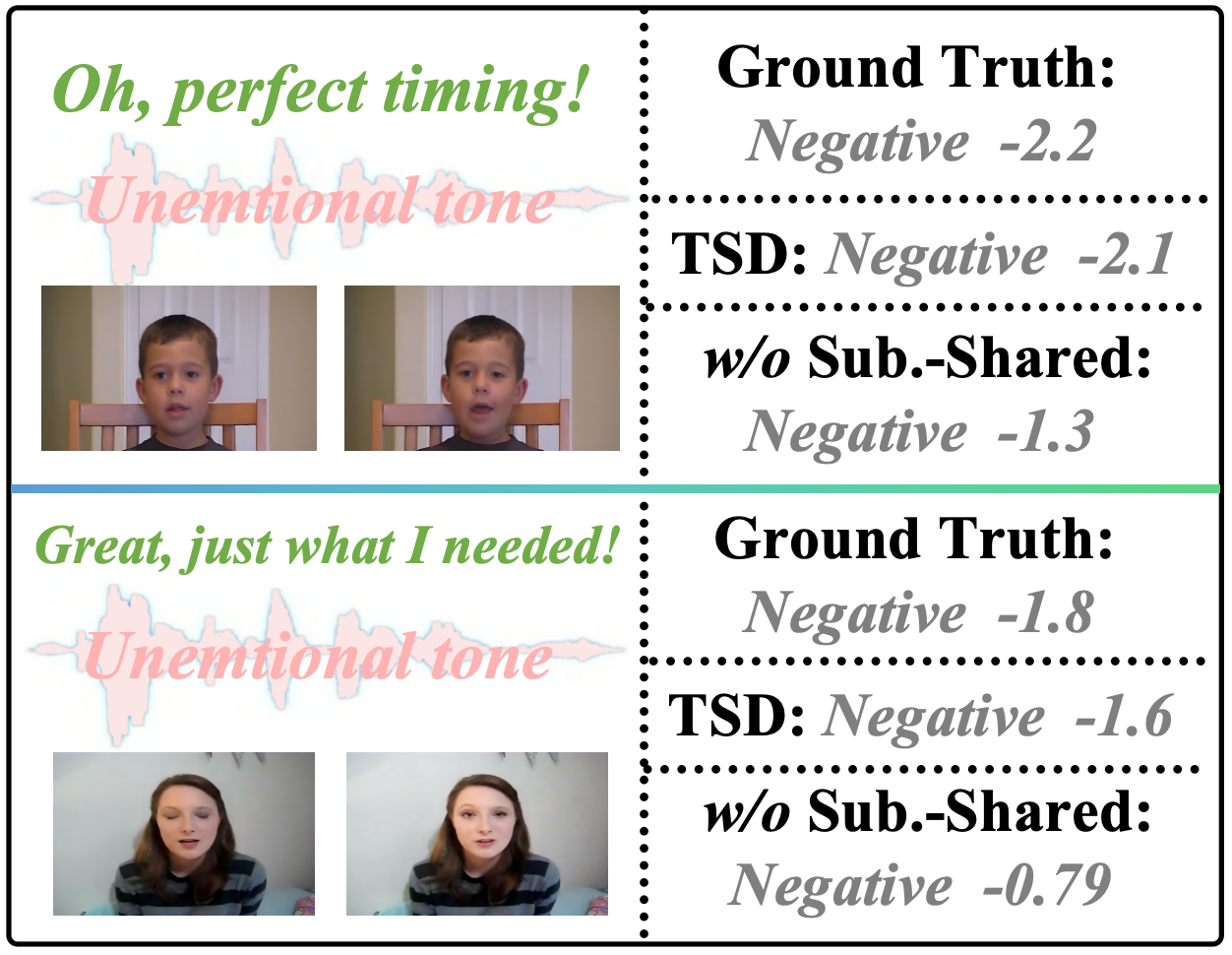}
    \caption{Qualitative examples illustrating that incorporating the submodally shared subspace enables TSD to better capture cross-modal cues (e.g., sarcasm), producing sentiment predictions closer to the ground truth.}
    \label{fig:qualitative}
\end{figure}

\begin{figure}[t]
    \centering
    \includegraphics[width=\linewidth]{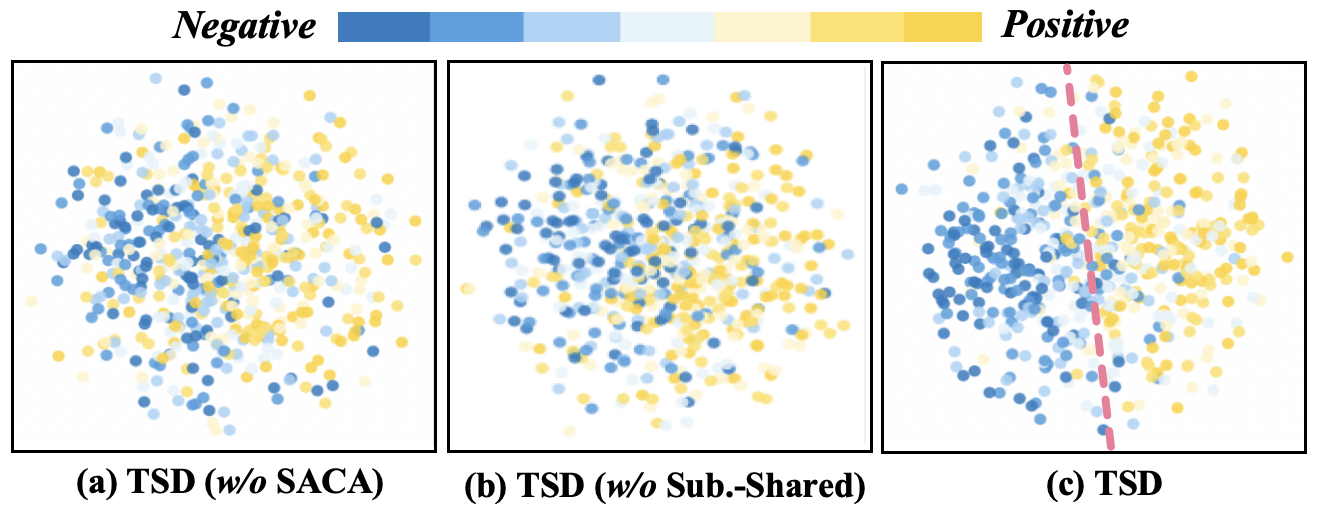}
    \caption{t-SNE visualization of feature distributions on CMU-MOSI.
    Colors indicate sentiment polarity from negative (dark) to positive (yellow).
    The full TSD model exhibits a clearer sentiment gradient than the variant without SACA and sub-shared subspaces.}
    \label{fig:tsne}
\end{figure}

\textbf{Visualization of feature distributions.}
We also visualize learned representations on CMU-MOSI using t-SNE~\cite{EMOE}, shown in Fig.~\ref{fig:tsne}.
The variant without SACA and Sub-Shared produces scattered, irregular distributions with weak sentiment structure.
In contrast, the full TSD model yields a more compact and continuous gradient from negative to positive sentiment.
Since MSA is a regression task, such gradient-like patterns are more desirable than discrete clusters, suggesting that TSD learns more coherent and semantically ordered representations.

\begin{figure}[t]
    \centering
    \includegraphics[width=\linewidth]{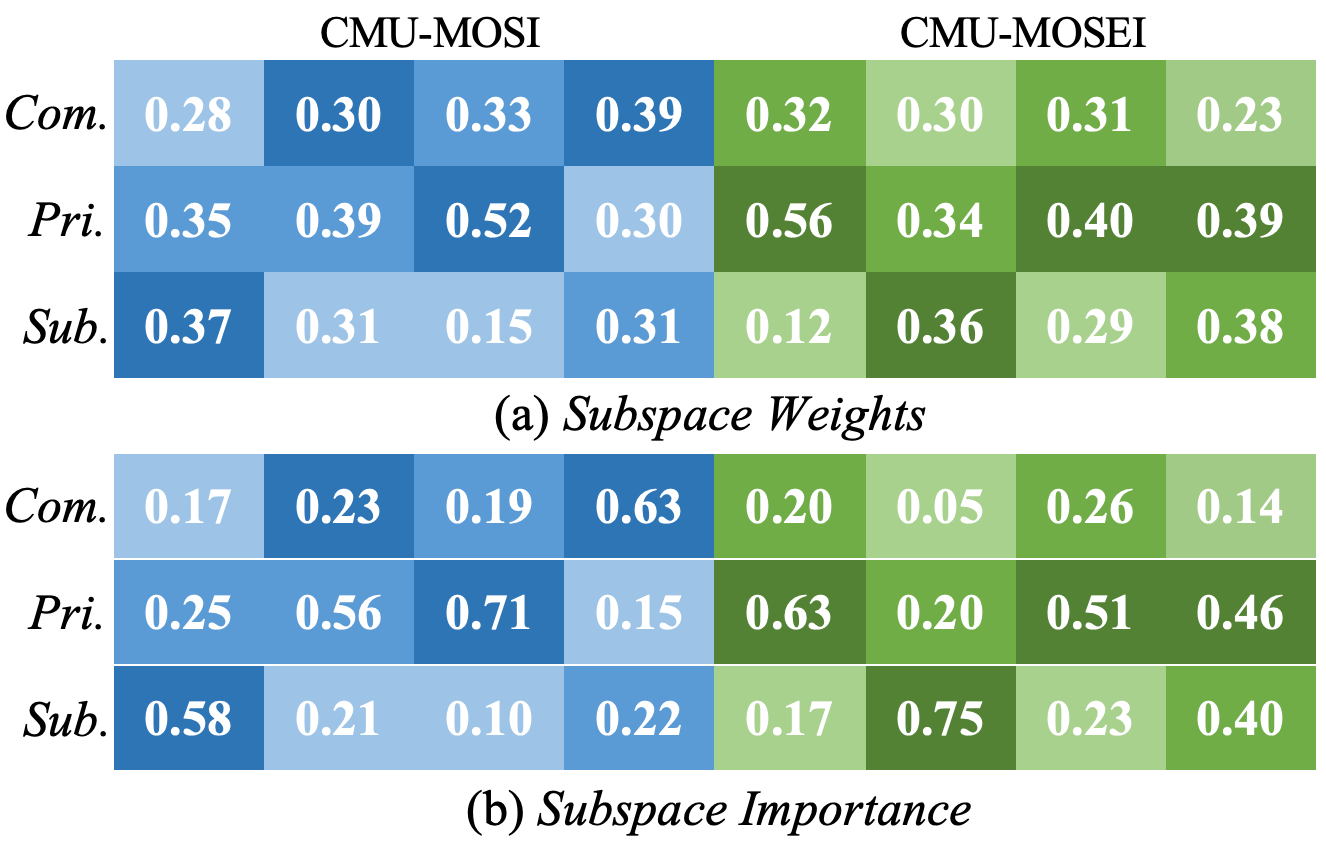}
    \caption{(a) Average fusion weights of each subspace (Common, Private, Sub-Shared) in TSD on CMU-MOSI and CMU-MOSEI.
    (b) Estimated contribution of each subspace during fusion.
    Higher values and darker colors denote greater subspace significance.}
    \label{fig:weight}
\end{figure}

\textbf{Visualization of subspace weights and importance.}
Finally, we analyze how TSD allocates attention across subspaces subspaces by visualizing learned fusion weights and estimated subspace contributions on MOSI and MOSEI (Fig.~\ref{fig:weight}).
On MOSI, the common and private subspaces receive higher average weights, whereas on MOSEI the weights are more balanced, reflecting the richer cross-modal patterns and the adaptive nature of SACA.
The contribution analysis further shows that TSD tends to rely on private subspaces when one modality is dominant, and on common or Sub-Shared subspaces when multimodal cues are complementary.
The strong agreement between learned weights and actual contributions supports both the effectiveness and interpretability of the tri-subspace design.

In addition, we conduct sensitivity analyses on key hyperparameters (e.g., subspace dimensionality and loss coefficients) and observe that TSD maintains strong performance over a wide range of settings, indicating robustness without heavy tuning. Full results, see Appendix.

%% file: sec/5_Conclusion.tex
\section{Conclusion}
This study presented the Tri-Subspace Disentanglement (TSD) framework to address submodally shared signals in multimodal sentiment analysis. TSD factorizes multimodal features into common, submodally shared, and private subspaces, and uses a decoupling supervisor with structured regularization to keep them distinct, enabling unified modeling of global, intermediate, and modality-specific cues. Built on these representations, a subspace-aware cross-attention module adaptively fuses information across subspaces. Experiments on MSA and MIR benchmarks show that TSD achieves state-of-the-art performance and transfers well to related tasks. In the future, we will expand TSD to a wider range of human-computer interaction scenarios.

%% file: sec/X_suppl.tex
\clearpage
\setcounter{page}{1}
\maketitlesupplementary

\appendix

\section{Model Zoo}

To evaluate the performance of our proposed Tri-Subspace Disentanglement (TSD) framework, we compare it with a selection of state-of-the-art methods in Multimodal Sentiment Analysis (MSA). We categorize these models into three main groups based on their approaches: traditional fusion methods, disentanglement-based methods, and recent fusion module designs.

\textbf{Traditional MSA Methods.} Early methods in multimodal sentiment analysis focus on directly combining features from different modalities. EF-LSTM~\cite{EMOE} is a recurrent model that uses LSTM networks for early fusion, integrating multimodal signals at the input level. TFN~\cite{TFN} applies tensor decomposition to model inter-modal interactions and integrate features from different modalities. LMF~\cite{LMF} adopts a low-rank matrix factorization technique to fuse multimodal information at an intermediate stage. These methods lay the foundation for multimodal fusion but are limited by their simplicity, often struggling to capture complex interactions between modalities, particularly in challenging tasks like sentiment analysis.

\textbf{Disentanglement-based Approaches.} More recent methods in multimodal sentiment analysis focus on disentangling modality-invariant and modality-specific components. This approach enables the model to better capture both shared and private features across different modalities. MISA~\cite{misa} introduces a shared-private disentanglement framework that aligns modality-specific features while preserving common features, aiming to improve cross-modal alignment. FDMER~\cite{FDMER} and ConFEDE~\cite{confede} propose similar disentanglement strategies but incorporate additional contrastive learning or fusion mechanisms to enhance alignment between modalities. DMD~\cite{dmd} and DLF~\cite{DLF} use distillation-based methods and contrastive learning to enhance multimodal fusion and reduce redundancy. 

\textbf{Recent Fusion Module Designs.} Recent advancements have introduced more sophisticated fusion mechanisms. These methods aim to model complex inter-modal interactions more effectively. For instance, MFN~\cite{EMOE} uses a memory-based fusion network that captures temporal relations across modalities, enabling the model to handle long-range dependencies. MuLT~\cite{MuLT} adopts a multimodal transformer architecture to explicitly model interactions between different modalities over time. Other models, such as EMOE~\cite{EMOE} and DEVA~\cite{DEVA}, use decision-level fusion and gating mechanisms to dynamically aggregate unimodal outputs.

\section{Statistical Significance Tests}
\label{appendix-sec:Statistical Significance Tests}

We assess the statistical significance of TSD by repeating all experiments with $n=5$ independent random seeds shared across methods under both aligned and unaligned settings on CMU-MOSI, CMU-MOSEI, and MIntRec. For each metric, we compute the mean and standard deviation and perform seed-wise paired $t$-tests comparing TSD against the strongest prior method on that metric within the same setting (as reported in Tables 1 and 2 in the main text.). $p$-values are adjusted for multiple comparisons within each table using Holm correction. The resulting statistics are summarized in Table~\ref{tab:significance_full}.

\textbf{Findings on MOSI/MOSEI.}
On both datasets and under both preprocessing regimes, TSD achieves statistically significant improvements on classification-oriented metrics (ACC$_2$/ACC$_7$/F1) after Holm correction, with all $p$-values at or below $0.05$.
The corresponding standard deviations across seeds are small (typically around $0.1$--$0.4$ percentage points for ACC/F1), indicating limited sensitivity to random initialization.
For MAE, the improvements are smaller in magnitude and several settings yield $p$-values slightly above $0.05$, which is consistent with the higher variability and lower effect size commonly observed on regression losses.
Overall, these results indicate that TSD brings the largest and most reliable gains on decision-centric metrics while remaining directionally favorable on MAE.

\textbf{Findings on MIntRec.}
On MIntRec, TSD improves Accuracy, F1, Precision, and Recall over the best prior method, with Holm-corrected $p$-values between $0.01$ and $0.04$ and modest standard deviations.
This shows that the proposed tri-subspace disentanglement and subspace-aware fusion transfer beyond sentiment regression/classification to multimodal intent recognition, supporting both accuracy- and discrimination-oriented metrics.

\textbf{Reproducibility.}
For all datasets, the Holm-corrected significance on classification metrics holds in both aligned and unaligned pipelines, showing that the conclusions are robust to the choice of preprocessing.
Taken together, TSD yields statistically reliable and empirically stable improvements across datasets, alignment regimes, and evaluation metrics of primary interest.

\begin{table}[t]
\centering
\setlength{\tabcolsep}{4pt}
\renewcommand{\arraystretch}{1.1}
\begin{tabular}{l||cc}
\toprule
\textbf{Metric} & \textbf{TSD} & \textbf{$p$-value (Holm)} \\
\midrule\midrule
\multicolumn{3}{c}{\it \textbf{CMU-MOSI (unaligned)}} \\
MAE ($\downarrow$)   & 0.691$\pm$0.005 & 0.08 \\
ACC$_7$ (\%)       & 49.0$\pm$0.3    & 0.006 \\
ACC$_2$ (\%)       & 86.5$\pm$0.1    & 0.001 \\
F1 (\%)            & 86.6$\pm$0.3    & 0.002 \\
\midrule
\midrule
\multicolumn{3}{c}{\it \textbf{CMU-MOSEI (unaligned)}} \\
MAE ($\downarrow$)   & 0.525$\pm$0.006 & 0.11 \\
ACC$_7$ (\%)       & 54.6$\pm$0.2    & 0.040 \\
ACC$_2$ (\%)       & 86.2$\pm$0.1    & 0.004 \\
F1 (\%)            & 86.2$\pm$0.4    & 0.020 \\
\midrule
\midrule
\multicolumn{3}{c}{\it \textbf{CMU-MOSI (aligned)}} \\
MAE ($\downarrow$)   & 0.701$\pm$0.005 & 0.040 \\
ACC$_7$ (\%)       & 48.8$\pm$0.3    & 0.010 \\
ACC$_2$ (\%)       & 86.3$\pm$0.1    & 0.001 \\
F1 (\%)            & 86.3$\pm$0.3    & 0.004 \\
\midrule
\midrule
\multicolumn{3}{c}{\it \textbf{CMU-MOSEI (aligned)}} \\
MAE ($\downarrow$)   & 0.529$\pm$0.005 & 0.090 \\
ACC$_7$ (\%)       & 54.9$\pm$0.2    & 0.025 \\
ACC$_2$ (\%)       & 85.8$\pm$0.1    & 0.006 \\
F1 (\%)            & 85.9$\pm$0.4    & 0.035 \\
\midrule
\midrule
\multicolumn{3}{c}{\it \textbf{MIntRec}} \\
Accuracy (\%)      & 73.67$\pm$0.50  & 0.040 \\
F1 (\%)            & 71.76$\pm$0.45  & 0.010 \\
Precision (\%)     & 72.86$\pm$0.45  & 0.020 \\
Recall (\%)        & 71.90$\pm$0.50  & 0.015 \\
\bottomrule
\end{tabular}
\caption{\textbf{Statistical significance of TSD.}
Means and standard deviations for TSD are computed over $n=5$ independent seeds.
$p$-values are obtained from seed-wise paired $t$-tests comparing TSD against the strongest prior method on the same metric and setting, followed by Holm correction for multiple comparisons within each block.}
\label{tab:significance_full}
\end{table}

\begin{figure*}[htbp]
    \centering
    \includegraphics[width=1\linewidth]{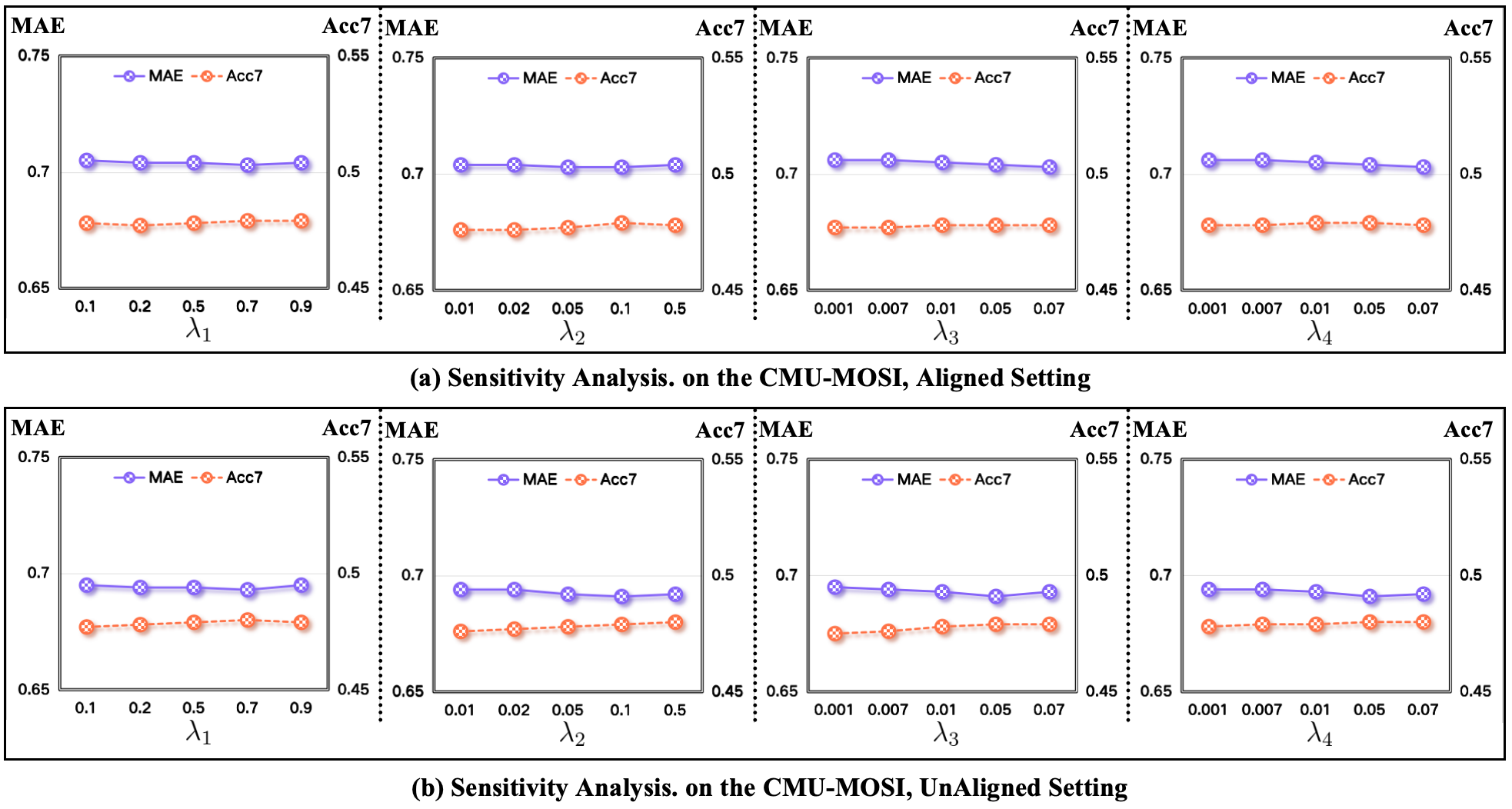}
    \caption{Sensitivity analysis of loss coefficients ($\lambda_1, \lambda_2, \lambda_3, \lambda_4$) on the CMU-MOSI dataset. Model performance (MAE and ACC$_7$) remains stable as each parameter is varied, illustrating the robustness of TSD to hyperparameter selection.}
    \label{fig:app-sen-mosi}
\end{figure*}

\begin{figure*}[htbp]
    \centering
    \includegraphics[width=1\linewidth]{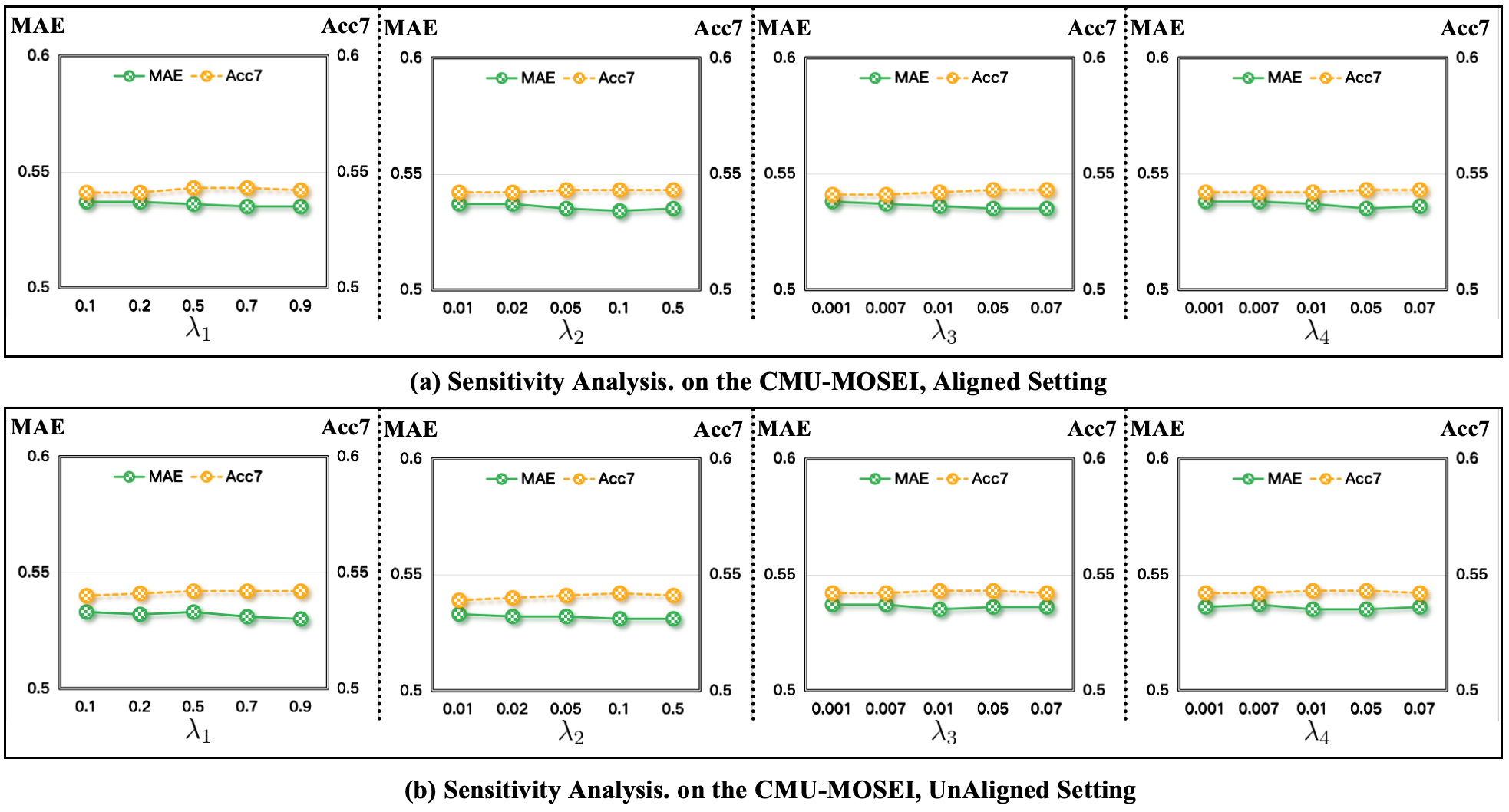}
    \caption{Sensitivity analysis of loss coefficients ($\lambda_1, \lambda_2, \lambda_3, \lambda_4$) on the CMU-MOSEI dataset. Model performance (MAE and ACC$_7$) remains stable as each parameter is varied, illustrating the robustness of TSD to hyperparameter selection.}
    \label{fig:app-sen-mosei}
\end{figure*}

\section{Sensitivity Analysis}
\label{app-sec:Sensitivity Analysis}

To assess the robustness of the proposed TSD framework, we conduct a sensitivity analysis of the primary loss coefficients ($\lambda_1, \lambda_2, \lambda_3, \lambda_4$) on both CMU-MOSI and CMU-MOSEI datasets. For each parameter, we vary its value across a reasonable range while keeping the others fixed to their default settings. The results, summarized in the Fig.~\ref{fig:app-sen-mosi} and Fig.~\ref{fig:app-sen-mosei}, indicate that the model’s performance in terms of MAE and ACC$_7$ remains stable under different parameter choices. No significant fluctuations or performance degradation are observed within the tested ranges, which demonstrates that TSD is not overly sensitive to the selection of these hyperparameters. This robustness confirms the reliability of our loss design and highlights the practical applicability of the proposed framework in real-world multimodal sentiment analysis scenarios.

%% file: main.bib
@String(ICASSP=	{ICASSP})

@String(AAAI = {AAAI})

@inproceedings{misa,
  title={Misa: Modality-invariant and-specific representations for multimodal sentiment analysis},
  author={Hazarika, Devamanyu and Zimmermann, Roger and Poria, Soujanya},
  booktitle={Proceedings of the 28th ACM international conference on multimedia},
  pages={1122--1131},
  year={2020}
}

@article{MInD,
  title={Mind: improving multimodal sentiment analysis via multimodal information disentanglement},
  author={Dai, Weichen and Li, Xingyu and Wang, Zeyu and Hu, Pengbo and Qi, Ji and Peng, Jianlin and Zhou, Yi},
  journal={arXiv preprint arXiv:2401.11818},
  year={2024}
}

@inproceedings{FDRL,
  title={Fine-grained disentangled representation learning for multimodal emotion recognition},
  author={Sun, Haoqin and Zhao, Shiwan and Wang, Xuechen and Zeng, Wenjia and Chen, Yong and Qin, Yong},
  booktitle={ICASSP 2024-2024 IEEE International Conference on Acoustics, Speech and Signal Processing (ICASSP)},
  pages={11051--11055},
  year={2024},
  organization={IEEE}
}

@inproceedings{FDMER,
  title={Disentangled representation learning for multimodal emotion recognition},
  author={Yang, Dingkang and Huang, Shuai and Kuang, Haopeng and Du, Yangtao and Zhang, Lihua},
  booktitle={Proceedings of the 30th ACM international conference on multimedia},
  pages={1642--1651},
  year={2022}
}

@inproceedings{dmd,
  title={Decoupled multimodal distilling for emotion recognition},
  author={Li, Yong and Wang, Yuanzhi and Cui, Zhen},
  booktitle={Proceedings of the IEEE/CVF conference on computer vision and pattern recognition},
  pages={6631--6640},
  year={2023}
}

@inproceedings{confede,
  title={Confede: Contrastive feature decomposition for multimodal sentiment analysis},
  author={Yang, Jiuding and Yu, Yakun and Niu, Di and Guo, Weidong and Xu, Yu},
  booktitle={Proceedings of the 61st Annual Meeting of the Association for Computational Linguistics},
  pages={7617--7630},
  year={2023}
}

@article{CGGM,
  title={Classifier-guided gradient modulation for enhanced multimodal learning},
  author={Guo, Zirun and Jin, Tao and Chen, Jingyuan and Zhao, Zhou},
  journal={Advances in Neural Information Processing Systems},
  volume={37},
  pages={133328--133344},
  year={2024}
}

@inproceedings{d2r,
  title={D2r: Dual-branch dynamic routing network for multimodal sentiment detection},
  author={Chen, Yifan and Li, Kuntao and Mai, Weixing and Wu, Qiaofeng and Xue, Yun and Li, Fenghuan},
  booktitle={Proceedings of the 2024 Conference on Empirical Methods in Natural Language Processing},
  pages={3536--3547},
  year={2024}
}

@article{Cmu-mosi,
  title={Multimodal sentiment intensity analysis in videos: Facial gestures and verbal messages},
  author={Zadeh, Amir and Zellers, Rowan and Pincus, Eli and Morency, Louis-Philippe},
  journal={IEEE Intelligent Systems},
  pages={82--88},
  year={2016}
}

@inproceedings{Cmu-mosei,
  title={Multimodal language analysis in the wild: Cmu-mosei dataset and interpretable dynamic fusion graph},
  author={Zadeh, AmirAli Bagher and Liang, Paul Pu and Poria, Soujanya and Cambria, Erik and Morency, Louis-Philippe},
  booktitle={Proceedings of the 56th Annual Meeting of the Association for Computational Linguistics},
  pages={2236--2246},
  year={2018}
}

@inproceedings{EMOE,
  title={EMOE: Modality-Specific Enhanced Dynamic Emotion Experts},
  author={Fang, Yiyang and Huang, Wenke and Wan, Guancheng and Su, Kehua and Ye, Mang},
  booktitle={Proceedings of the Computer Vision and Pattern Recognition Conference},
  pages={14314--14324},
  year={2025}
}

@inproceedings{DLF,
  title={DLF: Disentangled-language-focused multimodal sentiment analysis},
  author={Wang, Pan and Zhou, Qiang and Wu, Yawen and Chen, Tianlong and Hu, Jingtong},
  booktitle={Proceedings of the AAAI Conference on Artificial Intelligence},
  volume={39},
  pages={21180--21188},
  year={2025}
}

@inproceedings{DEVA,
  title={Enriching multimodal sentiment analysis through textual emotional descriptions of visual-audio content},
  author={Wu, Sheng and He, Dongxiao and Wang, Xiaobao and Wang, Longbiao and Dang, Jianwu},
  booktitle={Proceedings of the AAAI Conference on Artificial Intelligence},
  volume={39},
  pages={1601--1609},
  year={2025}
}

@inproceedings{Semi-IIN,
  title={Semi-IIN: Semi-supervised Intra-inter modal Interaction Learning Network for Multimodal Sentiment Analysis},
  author={Lin, Jinhao and Wang, Yifei and Xu, Yanwu and Liu, Qi},
  booktitle={Proceedings of the AAAI Conference on Artificial Intelligence},
  volume={39},
  pages={1411--1419},
  year={2025}
}

@inproceedings{MuLT,
  title={Multimodal transformer for unaligned multimodal language sequences},
  author={Tsai, Yao-Hung Hubert and Bai, Shaojie and Liang, Paul Pu and Kolter, J Zico and Morency, Louis-Philippe and Salakhutdinov, Ruslan},
  booktitle={Proceedings of the conference. Association for computational linguistics. Meeting},
  pages={6558},
  year={2019}
}

@inproceedings{TFN,
    title = {Tensor Fusion Network for Multimodal Sentiment Analysis},
    author = {Zadeh, Amir  and Chen, Minghai  and
      Poria, Soujanya  and
      Cambria, Erik  and
      Morency, Louis-Philippe},
    booktitle = {Proceedings of the 2017 Conference on Empirical Methods in Natural Language Processing},
    pages = {1103--1114}, 
  year = {2017}
}

@article{LMF,
  title={Efficient low-rank multimodal fusion with modality-specific factors},
  author={Liu, Zhun and Shen, Ying and Lakshminarasimhan, Varun Bharadhwaj and Liang, Paul Pu and Zadeh, Amir and Morency, Louis-Philippe},
  journal={arXiv preprint arXiv:1806.00064},
  year={2018}
}

@article{GELU,
  title={Gaussian error linear units (gelus)},
  author={Hendrycks, Dan and Gimpel, Kevin},
  journal={arXiv preprint arXiv:1606.08415},
  year={2016}
}

@inproceedings{self-mm,
  title={Learning modality-specific representations with self-supervised multi-task learning for multimodal sentiment analysis},
  author={Yu, Wenmeng and Xu, Hua and Yuan, Ziqi and Wu, Jiele},
  booktitle={Proceedings of the AAAI conference on artificial intelligence},
  pages={10790--10797},
  year={2021}
}

@article{ALMT,
  title={Learning language-guided adaptive hyper-modality representation for multimodal sentiment analysis},
  author={Zhang, Haoyu and Wang, Yu and Yin, Guanghao and Liu, Kejun and Liu, Yuanyuan and Yu, Tianshu},
  journal={arXiv preprint arXiv:2310.05804},
  year={2023}
}

@inproceedings{MIntRec,
author = {Zhang, Hanlei and Xu, Hua and Wang, Xin and Zhou, Qianrui and Zhao, Shaojie and Teng, Jiayan},
title = {MIntRec: A New Dataset for Multimodal Intent Recognition},
booktitle = {Proceedings of the 30th ACM International Conference on Multimedia},
pages = {1688–1697},
numpages = {10},
year = {2022}
}

@inproceedings{MAG-BERT,
  author       = {Wasifur Rahman and Md. Kamrul Hasan and Sangwu Lee and AmirAli Bagher Zadeh and Chengfeng Mao and Louis{-}Philippe Morency and Mohammed E. Hoque},
  title        = {Integrating Multimodal Information in Large Pretrained Transformers},
  booktitle    = {Proceedings of the 58th Annual Meeting of the Association for Computational Linguistics},
  pages        = {2359--2369},
  year         = {2020}
}

@article{cta-net,
  title={CTA-Net: A CNN-Transformer Aggregation Network for Improving Multi-Scale Feature Extraction},
  author={Meng, Chunlei and Yang, Jiacheng and Lin, Wei and Liu, Bowen and Zhang, Hongda and Gan, Zhongxue and others},
  journal={arXiv preprint arXiv:2410.11428},
  year={2024}
}

@ARTICLE{rts-vit,
  author={Meng, Chunlei and Lin, Wei and Liu, Bowen and Zhang, Hongda and Gan, Zhongxue and Ouyang, Chun},
  journal={IEEE Journal of Biomedical and Health Informatics}, 
  title={RTS-ViT: Real-Time Share Vision Transformer for Image Classification}, 
  year={2025},
  volume={29},
  number={5},
  pages={3576-3586},
 }

@inproceedings{HSIC,
  title={Measuring statistical dependence with Hilbert-Schmidt norms},
  author={Gretton, Arthur and Bousquet, Olivier and Smola, Alex and Sch{\"o}lkopf, Bernhard},
  booktitle={International conference on algorithmic learning theory},
  pages={63--77},
  year={2005},
  organization={Springer}
}

@inproceedings{GsiT,
    title ={Multimodal Transformers are Hierarchical Modal-wise Heterogeneous Graphs},
    author = {in, Yijie  and Peng, Junjie  and Lin, Xuanchao  and Yuan, Haochen  and Wang, Lan  and Zheng, Cangzhi},
    booktitle = {Proceedings of the 63rd Annual Meeting of the Association for Computational Linguistics (Volume 1: Long Papers)},
    year = 2025
}

@inproceedings{CAGC,
  title={Contextual augmented global contrast for multimodal intent recognition},
  author={Sun, Kaili and Xie, Zhiwen and Ye, Mang and Zhang, Huyin},
  booktitle={Proceedings of the IEEE/CVF Conference on Computer Vision and Pattern Recognition},
  pages={26963--26973},
  year={2024}
}

@article{TSDA,
      title={Temporal-Spatial Decouple before Act: Disentangled Representation Learning for Multimodal Sentiment Analysis}, 
      author={Chunlei Meng and Ziyang Zhou and Lucas He and Xiaojing Du and Chun Ouyang and Zhongxue Gan},
      journal={arXiv preprint arXiv:2601.13659},
      year={2026}
}

@inproceedings{CF-ViT,
  author={Meng, Chunlei and Lin, Wei and Yang, Jiacheng and Liu, Yi and Zhang, Hongda and Chen, Yuning and Liu, Bowen and Zhou, Ziqin and Ouyang, Chun and Gan, Zhongxue and Wu, Dunzhao and Nie, Zhihua},
  booktitle={2025 IEEE International Conference on Systems, Man, and Cybernetics (SMC)}, 
  title={CF-ViT: Cross-Feature Vision Transformer for Improving Feature Learning on Tiny Datasets}, 
  year={2025},
  pages={6919-6926}
}

@article{qiao,
  title={We-math: Does your large multimodal model achieve human-like mathematical reasoning?},
  author={Qiao, Runqi and Tan, Qiuna and Dong, Guanting and Wu, Minhui and Sun, Chong and Song, Xiaoshuai and GongQue, Zhuoma and Lei, Shanglin and Wei, Zhe and Zhang, Miaoxuan and others},
  journal={arXiv preprint arXiv:2407.01284},
  year={2024}
}

@article{MGJR,
title = {Multi-grained teacher–student joint representation learning for surface defect classification},
journal = {Journal of Industrial Information Integration},
author = {Chunlei Meng and Jiacheng Yang and Wei Lin and Linqiang Hu and Bowen Liu and Zhuo Zou and LiDa Xu and Zhongxue Gan and Chun Ouyang},
volume = {48},
pages = {100958},
year = {2025}
}
